\documentclass{article}
\usepackage[utf8]{inputenc}
\usepackage[affil-it]{authblk}
\usepackage{graphicx}
\usepackage[margin=1in]{geometry}
\usepackage{cite}
\title{Molecular Dynamics-Based Strength Estimates of Beta-Solenoid Proteins}

\author{Amanda S. Parker, Krishnakumar M. Ravikumar, Daniel L. Cox*}
\affil{Department of Physics, University of California, Davis, California \\
*Corresponding author e-mail address: cox@physics.ucdavis.edu}

\begin{document}

\maketitle

\section{Abstract}

The use of beta-solenoid proteins as functionalizable, nanoscale, self-assembling molecular building blocks may have many applications, including templating the growth of wires or higher-dimensional structures. By understanding their mechanical strengths, we can efficiently design the proteins for specific functions. We present a study of the mechanical properties of seven beta-solenoid proteins using GROMACS molecular dynamics software to produce force/torque-displacement data, implement umbrella sampling of bending/twisting trajectories, produce Potentials of Mean Force (PMFs), extract effective spring constants, and calculate rigidities for two bending and two twisting directions for each protein. We examine the differences between computing the strength values from force/torque-displacement data alone and PMF data, and show how higher precision estimates can be obtained from the former. In addition to the analysis of the methods, we report estimates for the bend/twist persistence lengths for each protein, which range from  0.5-3.4 $\mu$m. We note that beta-solenoid proteins with internal disulfide bridges do not enjoy enhanced bending or twisting strength, and that the strongest correlate with bend/twist rigidity is the number of hydrogen bonds per turn. In addition, we compute estimates of the Young's modulus ($Y$) for each protein, which range from $Y$ = 3.5 to 7.2 GPa.

\section{Introduction}

To realize the promise of bottom-up self-assembly on an industrial scale it is necessary to identify programmable, environmentally robust, mechanically strong, and highly functionalizable building blocks at a molecular level that offer industrial scalability. DNA is highly programmable and functionalizable\cite{dnapro}, but lacks environmental toughness, mechanical rigidity, and possibly industrial scalability\cite{dnacon}. Proteins such as S-layers\cite{slayers} and synthetic spider silk\cite{spidersilk} are potentially scalable through microbial expression, environmentally robust, and mechanically strong, but are less programmable and functionalizable because the sequence-to-structure matching is not precisely known.  

We have recently argued that modified beta solenoid proteins which have similar mechanical strength to spider silk can bridge the gap between programmability and functionalizability on the one side and scalability and robustness on the other\cite{peralta_2015}. In particular, while they share the amyloid character with the key component of spider silk, the sequence-to-structure map is fully known, so they can be highly engineered in terms of modifying the side chains for binding biomolecules, templating or capturing nanoparticles, or engendering lateral self-assembly.  They also have controllable length and in some cases controllable twist.  

In this paper we carry out a comprehensive study of the twist and bend rigidities of all seven candidate modified beta-solenoid proteins using molecular dynamics simulations with umbrella sampling. In a previous study, we considered bend and twist mechanics for four beta solenoids using umbrella sampling and weighted histogram analysis methodology to generate potentials of mean force (PMF) as a function of bend deflection or twist angle\cite{heinz_2015}.  In this work, besides the additional proteins, we (i) establish the greater precision of regression analysis on binned force or torque vs. displacement data for the umbrella sampling than from PMF curves, (ii) show that internal disulfide bridges do not enhance the bend or twist strength (in fact, they may weaken it in once case), (iii) determine the Young's moduli for the proteins, and (iv) determine that the strongest correlate with bending or twist strength is the number of hydrogen bonds per layer, although the internal volume packing fraction is also important. 

\section{Methods} \label{methods}

\subsection{Proteins}

We studied the mechanical properties of seven beta-solenoid proteins: an antifreeze protein from the \textit{Rhagium inquisitor} longhorned beetle (RiAFP) \cite{riafp_pdb}, an antifreeze protein from the \textit{tenebrio molitor} beetle (TMAFP-m1) \cite{1ezg_tmafp_pdb}, a wild-type spruce budworm antifreeze protein (SBAFP) \cite{1m8n_wtsbafp_pdb}, a modified spruce budworm antifreeze protein (SBAFP-m1) \cite{peralta_2015}, a modified pentapeptide repeat protein (SQBSP-m1) \cite{sqbsp}, a modified rye-grass antifreeze protein (RGAFP-m1) \cite{peralta_2015}, and a modified catalytic protein of E. coli (TRBSP-m1) \cite{trbsp}. 

To produce the modified TMAFP-m1, we removed the first turn of the wild-type protein. The SBAFP is identical to the wild-type, except for the removal of a small hook on its end. The SBAFP-m1 further differs from the SBAFP by mutating several residues, including the cysteines which create disulfide bonds in the wild-type protein. The end caps of SQBSP were removed to produce our SQBSP-m1. Our RGAFP-m1 differs from the wild-type by mutated residues, particularly at the interfaces to aid in binding. Similar to the modification of the TMAFP, the TRBSP was modified by having its non-beta-sheet first turn removed to produce the modified TRBSP-m1. For protein sequences, see Supporting Information.

\begin{figure}
	\center
	\includegraphics[width=0.8\textwidth]{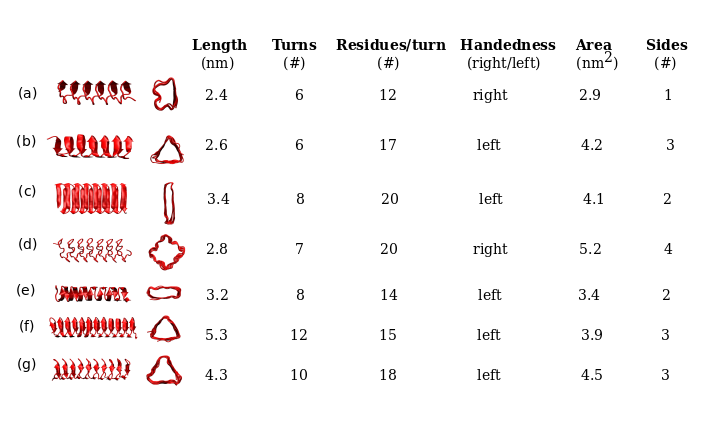}
	\caption{Images and characteristics of all seven proteins of study. (a) TMAFP-m1, (b) SBAFP, (c) RiAFP, (d) SQBSP-m1, (e) RGAFP-m1, (f) SBAFP-m1, (g) TRBSP-m1. First, long and cross-sectional views, not to scale, are shown. Then length, total number of helical turns, number of amino acids per turn, handedness, cross-sectional area, and the number of sides per helical turn for each protein are reported. The length of each protein was estimated as the average distance between the center of masses of its first and last turns, over a 50 ps simulation. To determine the cross-sectional area of each protein, we used a Monte Carlo sampling approach. (See Supporting Information for more details.) Protein images were made using VMD \cite{humphrey_96} and the Tachyon library \cite{stone_98}.}
	\label{proteins_details}
\end{figure}

\begin{figure}
	\center
	\includegraphics[width=0.5\textwidth]{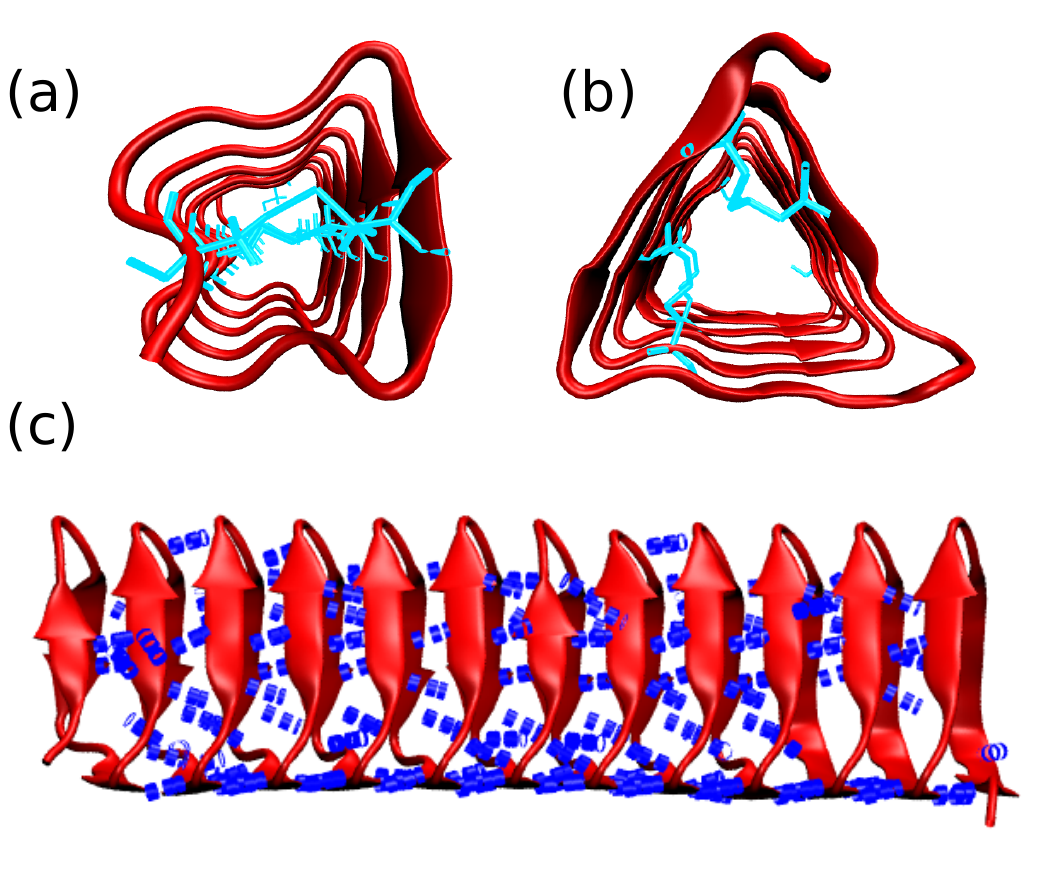}
	\caption{Detail of intralayer cysteine disulfide bonds in (a) TMAFP-m1 and (b) SBAFP. (c) Illustration of interlayer hydrogen bonds, here shown in SBAFP-m1.}
	\label{bonds}
\end{figure}

These proteins vary in number of sides, cross-sectional areas, lengths, number of total turns, and number of total amino acids (Fig. \ref{proteins_details}). Adjacent turns of each protein are connected by hydrogen bonds (Fig. \ref{bonds} (c)). In TMAFP-m1 and SBAFP, there are also intralayer cysteine disulfide bonds (Fig. \ref{bonds} (a), (b)).

The bending and torsional properties of SBAFP-m1, SQBSP-m1, RGAFP-m1, and TRBSP-m1 have been studied previously \cite{heinz_2015}, and the tensile strength of RGAFP-m1 has also been studied \cite{solar_2014}. In this paper we extend and further validate that work in addition to studying the RiAFP, TMAFP-m1 and SBAFP.

\subsection{Pulling simulations}

GROMACS 4.6.5 \cite{gromacs}, a molecular dynamics (MD) package, was used to perform all simulations. These simulations modeled the strength of a selection of beta-solenoid proteins under the influence of external forces by allowing us to measure the forces and displacements at breaking and thus calculate their flexural and torsional rigidities, persistence lengths, and Young's moduli. A description of our simulation procedure is as follows. 

We used the CHARMM27 \cite{charmm} force field and the TIP3P \cite{tip3p} water model for explicit solvent.

The protein was placed in a box and surrounded by solvent (water) molecules. The box was built with periodic boundary conditions, but such that each atom was no closer than 1 nm from any box wall. If necessary, the system was charge-neutralized by adding chlorine or sodium atoms to the solvent. 

We then relaxed the system through a series of equilibration simulations. These were done in two sets. During the first set, the backbone atoms were constrained using springs, each of force constant 1000 kJ/(mole-nm$^{2}$), and under these conditions the following five simulations were performed: 1. The energy of the system was minimized. 2. The system was heated to T = 300 K over a 50 ps period, using a Berendsen temperature coupling scheme \cite{temp_coupling}. 3. The system was allowed to relax at constant temperature and volume (an NVT simulation) during a 30 ps period, also using Berendsen thermostat. 4. The system was relaxed at constant temperature and pressure (an NPT simulation) during a 50 ps period, using Parrinello-Rahman pressure coupling \cite{barostat}. 5. The constraints on the backbone were removed and another NPT simulation was done in order to confirm the stability of the protein structure. The solvent was then deleted and the protein was aligned so that its helical axis lay along the x-axis. Then the protein was resolvated and the second set of equilibration simulations were performed. The steps for this set were identical to the first one, except that all of the C-alpha atoms, as opposed to the backbone atoms, were constrained throughout.

Finally the protein was subjected to force-pulling. The center of mass of the C-alpha atoms of the first turn of the protein was strongly constrained (essentially fixed) using a spring with force constant 10000 kJ/(mole-nm$^{2}$), and the center of mass of the C-alpha atoms of the last turn of the protein was pulled in the direction of interest. This pulling was done by attaching a spring and moving it at a constant speed of 1 nm/ns (or for twisting, 0.09 degrees/ps). The protein was pulled over a period of 1 ns to 2 ns, during which the secondary structure of the protein was broken. Because of the high speeds simulated during force-pulling, we refer to the pulling simulations as non-equilibrium simulations.

\begin{figure}
	\center
	\includegraphics[width=0.5\textwidth]{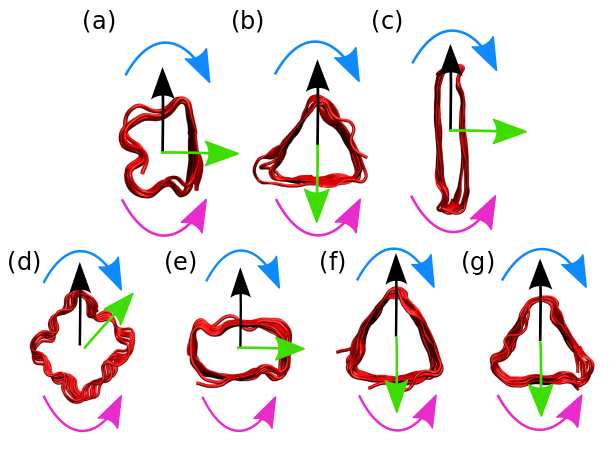}
	\caption{All bending and twisting directions, superimposed onto the protein cross-sections, are shown. The letters correspond to those in Fig. \ref{proteins_details}.}
	\label{directions}
\end{figure}

There were five pulling types of interest for each protein-- bending along two directions perpendicular to the helical axis, twisting with and against the helical twist, and stretching parallel to the helical axis of the protein. The bending and twisting directions are shown in Fig. \ref{directions}. Each protein is shown in cross-section, looking down its helical axis. The black and green arrows represent the two bending directions and the blue and magenta arrows represent the two twisting directions. These directions were chosen based on the protein geometries.

\subsection{Validation}

In order to expand on previous work \cite{heinz_2015,solar_2014} and to validate the methodology, three non-equilibrium pulling simulations were done for each direction, with the same initial parameters, in order to produce a distribution of data sets for analysis.

In addition, for the bending and twisting simulations, umbrella sampling \cite{umbrella} was done on the trajectory of each pulling simulation. By this we mean that 10-20 sample configurations, evenly-spaced along the reaction coordinate, were taken from each pulling simulation trajectory, and on each of these configurations, an equilibrium simulation was performed. This allowed us to sample the equilibrated configuration space of the protein more fully, without having to perform the pulling simulation itself in equilibrium (necessitating an infinitesimally small pulling speed). During these equilibrium simulations, the center of mass of the first turn was again strongly constrained, and the last turn was also lightly constrained. Each simulation lasted for 1-2 ns. 

Using the Weighted Histogram Analysis Method (WHAM) \cite{wham}, we then constructed from these equilibrium simulations a potential of mean force (PMF) for the protein along the pulling direction. To enable the WHAM analysis, the umbrella sampling configurations were also selected for reasonable overlap of the histograms of conformations along the bend or twist coordinate.  

\begin{figure}
	\center
	\includegraphics[width=0.4\textwidth]{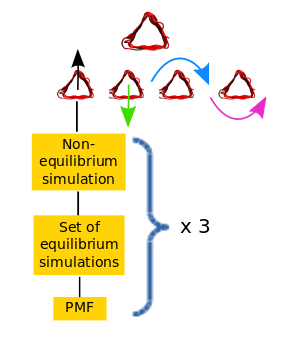}
	\caption{A schematic diagram of the simulation procedure. For a particular protein, multiple pulling directions of interest are studied, indicated by arrows. For each direction, three pulling simulations are done. For each pulling simulation, one set of umbrella simulations are done. From each of these umbrella simulations, a PMF is produced. Only the simulations for one pulling direction for one protein are shown, but the procedure is similar for all directions and all proteins.}
	\label{procedure}
\end{figure}

Using umbrella sampling followed by WHAM to construct PMFs has been frequently employed before \cite{roux95}, despite known issues with quantifying the error incurred through these techniques \cite{zhu12}. One way around this is to use the output from the umbrella sampling simulations directly. During a sample simulation, the force (or torque) and position of the protein each fluctuate around a mean value. 

As we expect, the force as a function of displacement before breakage is linear for both the direct non-equilibrium simulation data and average values from the equilibrium sample simulations. Therefore we fit lines to these data sets and extracted effective spring constants (taken to be the slopes of the force-displacement data). Quadratics were fit to the PMFs and spring constants were extracted from these as twice the coefficient of the quadratic term. These spring constants were then used to calculate strength quantities.

We also note that during the non-equilibrium and equilibrium simulations, the LINCS constraint algorithm \cite{hess_1997} was used, and we found that when we used this algorithm for all the bonds in the protein, the spread in results was larger than when we used it on only the hydrogen bonds. Therefore only hydrogen bonds were constrained using the LINCS algorithm. This choice is consistent with the recommendation in the GROMACS documentation for use of LINCS with the CHARMM force fields \cite{vanderspoel_2013}.

\section{Results and discussion} \label{results}

\subsection{Persistence lengths}

As described in Methods, each protein was subjected to pulling in five different directions, which were chosen based on the symmetry of the protein's secondary structure. Two of these directions were perpendicular to the helical axis and are therefore referred to as the ``bending'' directions, and two were rotations about the helical axis and are referred to as the ``twisting'' directions. In each of these cases, one end of the protein was fixed, while the other end was attached to a stiff spring moving at a constant speed in the pulling direction. At some point during each simulation, the protein's secondary structure broke, rupturing the hydrogen bonds that connect adjacent turns. The behavior of the force/torque and displacement of the protein before this breakage occurred was analyzed to ultimately provide an effective spring constant for each protein, in each pulling direction, from which flexural or torsional rigidities and persistence lengths, were calculated. The flexural rigidity was calculated as EI = (1/3)$k_{i}L^3$ and the torsional rigidity was calculated as GJ = $k_{j}L$, where $i$ is a bending direction and $j$ is a twisting direction. The persistence length is just the rigidity divided by $k_{b}T$, where $k_{b}$ is the Boltzmann constant and $T$=300 K. Our focus was therefore on the analysis of the non-equilibrium simulation data sets to produce spring constants.

First we analyzed the force and displacement of each protein, over each non-equilibrium simulation, directly. In all cases we saw linear behavior of the force (torque) vs. displacement (angular displacement) data before breakage. In other words the force (torque) increased as the displacement (angle) in the pulling direction increased, until the breaking force (torque) was reached, at which point the secondary structure of the protein was compromised. This force (torque) vs. displacement (angular displacement) behavior was linear and fitting a line ($kx+b$) to it provided an estimate for a spring constant $k$.

Second, we sampled the non-equilibrium simulation trajectory and ran equilibrium simulations on each of the chosen umbrella sampling configurations. The average force (torque) and displacement (angular displacement) value during each sample equilibrium simulation was plotted, along with their standard deviations, and a line was fit to this data, providing another estimate of the spring constant $k$.

Finally, using the sample equilibrium simulations, we used WHAM to produce a PMF for each non-equilibrium simulation. To this we fit a quadratic form ((1/2)$k(x-x_0)^{2}$ for bending displacement $x$ with a similar quadratic form in angle for the torsion) to determine the spring constant. 

Each non-equilibrium pulling simulation was repeated three times, enabling us to find three spring constants for each of the methods described above, which also revealed the variation in results in each method. Using the average and standard deviation of the spring constant from each method, we calculated flexural and torsional persistence lengths for each protein, in each pulling direction.

The validity of using a simple beam model to calculate values for the flexural and torsional rigidities of the proteins has been shown previously, particularly for the RGAFP-m1, SQBSP-m1, and TRBSP-m1 \cite{heinz_2015}, and we confirmed it using similar methods (see Supporting Information).

\begin{figure}
	\center
	\includegraphics[width=0.5\textwidth]{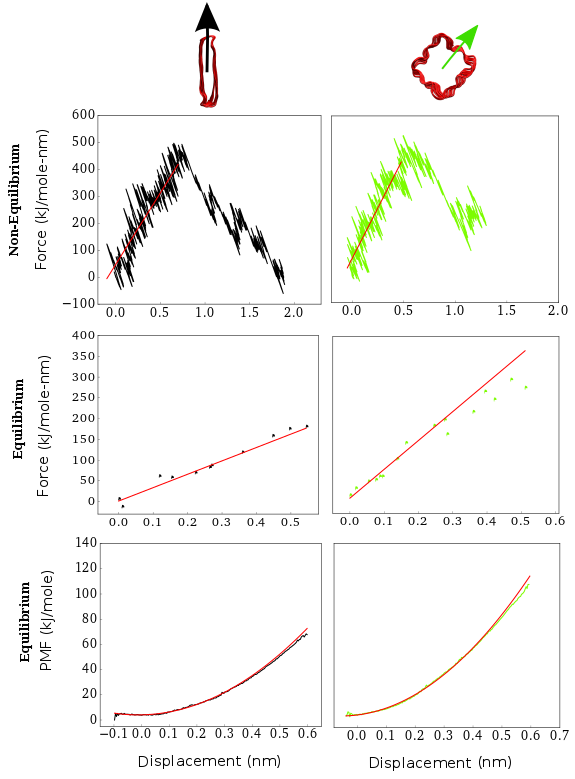}
	\caption{Two examples of results from non-equilibrium and equilibrium bending simulations. Each column represents data from one simulation. The protein at the top of each column was pulled in the direction indicated by the arrow shown. The first row shows fits to the linear regions of the non-equilibrium simulation data. The second row shows fits to the mean force and position values from the equilibrium simulations sampled from the non-equilibrium pulling simulation. The standard errors are shown, but are smaller than the marker size. The third row shows fits to the PMFs, which were produced using WHAM and the data from the sample simulations.}
	\label{example_data_bend}
\end{figure}

Examples of typical behavior in the force-displacement (from both non-equilibrium and equilibrium simulations) and PMF curves is shown in Fig. \ref{example_data_bend}. The left column in the figure shows data for bending the RiAFP in the direction indicated by the black arrow. The right column shows data for bending the SQBSP-m1 in the direction indicated by the green arrow. The first row shows force-displacement curves, which resulted from the non-equilibrium pulling simulations. In red, we have plotted the fit line to the region before the protein breaks. The second row shows average force-displacement data from the equilibrium sample simulations before breakage, with fit lines in red. The third row shows potential energy (PMF) curves, produced through equilibrium simulations and WHAM, with the quadratic fits in red. This data represents only one simulation for each protein in each of the directions indicated, but the qualitative results are typical of all simulations, for all proteins, in all directions.

\begin{figure}
	\center
	\includegraphics[width=0.5\textwidth]{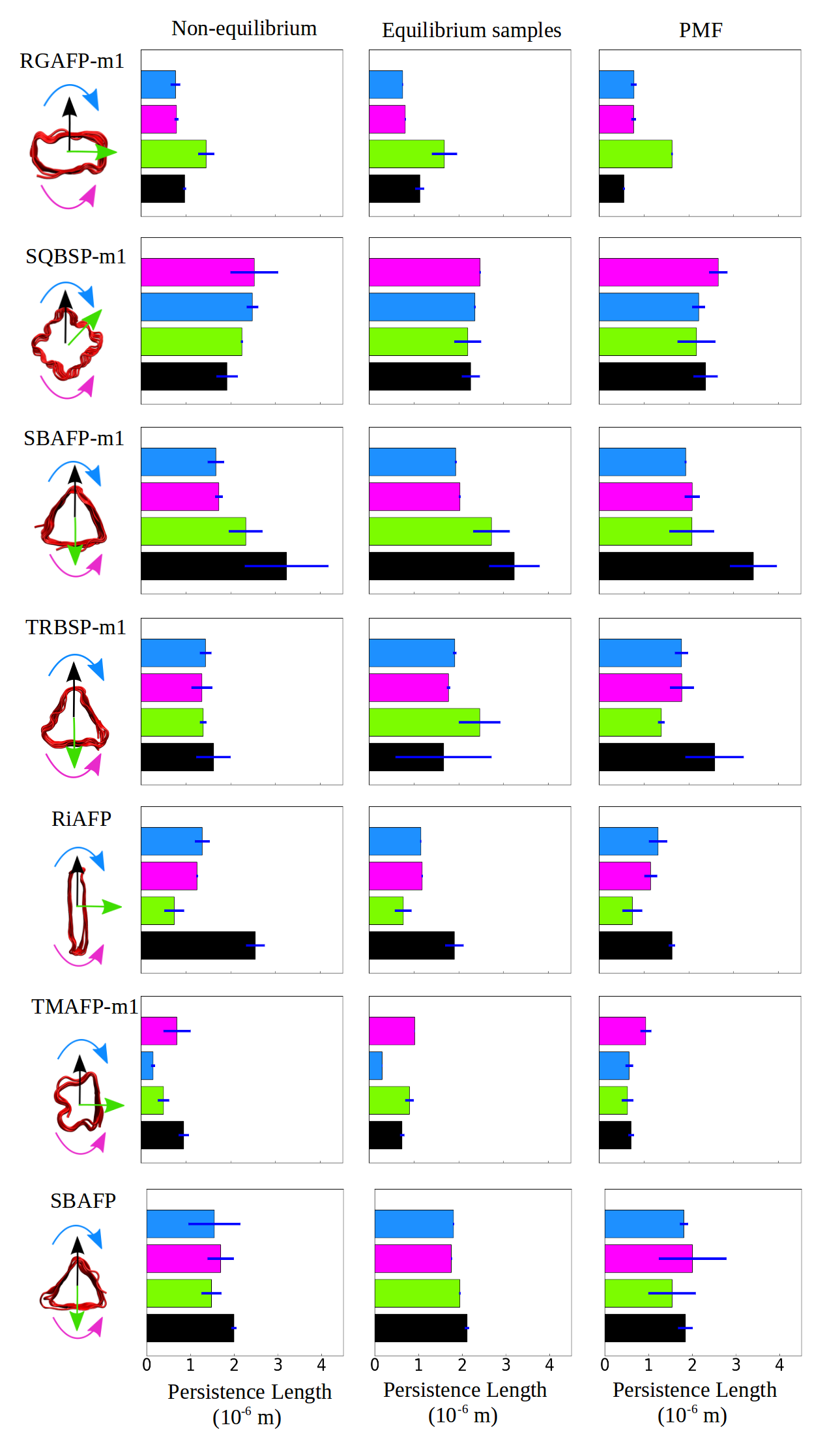}
	\caption{Flexural and torsional persistence lengths for each protein and each pulling direction. The first column represents values derived from fits to the non-equilibrium pulling simulations alone. The second column represents data from fits to the force/torque and displacement data from the equilibrium sample simulations. The third column represents data from fits to the PMFs. Each value and error is calculated using the average and standard deviation of the three spring constants obtained in each method.}
	\label{peristence_lengths_bar}
\end{figure}

The flexural persistence lengths, calculated using the three methods described above, are shown in Fig. \ref{peristence_lengths_bar}.

\begin{figure}
	\center
	\includegraphics[width=0.5\textwidth]{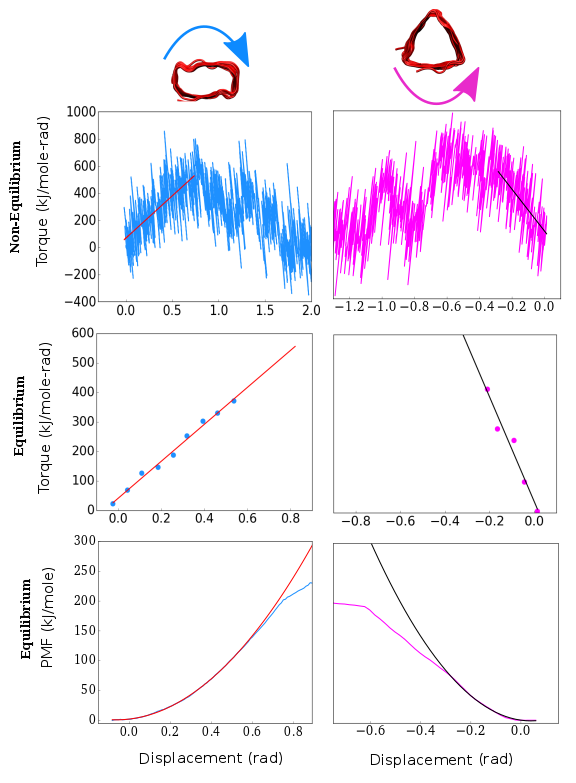}
	\caption{Two examples of results from non-equilibrium and equilibrium twisting simulations. Each column represents data from one simulation. The protein at the top of each column was rotated in the direction indicated by the arrow shown. The first row shows fits to the linear regions of the non-equilibrium simulation data. The second row shows fits to the mean torque and angular position values from the equilibrium simulations sampled from the non-equilibrium pulling simulation. The standard errors are shown, but are smaller than the marker size. The third row shows fits to the PMFs, which were produced using WHAM and the data from the sample simulations.}
	\label{example_data_twist}
\end{figure}

Analogous to the bending simulations, twisting simulations were conducted on each protein with and against their natural twists. Examples of typical behavior for these simulations, as well as for the equilibrium simulations and PMF curves, are shown in Fig. \ref{example_data_twist}.

The torsional persistence lengths, calculated using the three methods described above, are also shown in Fig. \ref{peristence_lengths_bar}.

Our results show that the persistence lengths determined by fitting quadratics to PMFs or by fitting a line to the forces/torques and displacements of sample simulations give similar average values, but that the standard deviations are in general smaller for the linear fits to the samples. This difference is even more pronounced if we assume the errors on the PMF data are lower bounds, given that the PMF itself comes with difficult-to-quantify errors from the WHAM algorithm. Therefore, if the PMF is not necessary, we suggest using the data from the umbrella simulations alone to extract quantities like the effective spring constant. The fits to the non-equilibrium force/torque-displacement data produces the largest standard deviations, but the average values from this method are comparable to the other two. We note that if qualitative estimates of the average spring constants are desired, forgoing umbrella sampling and PMF production would still yield useful results.

We note that the SBAFP-m1 has relatively large persistence lengths in all pulling directions, and the largest in the bending directions, while the TMAFP-m1 has the smallest. The SBAFP has twisting persistence lengths comparable to the SBAFP-m1, but bending persistence lengths that are slightly smaller, despite the intramolecular cysteine bonds found in the SBAFP (but not the SBAFP-m1). 

Upon further investigation, we found that the inner cysteine-cysteine bonds found in the SBAFP and TMAFP-m1 do not increase the bending or torsional rigidity of the proteins. The persistence lengths for bending the SBAFP and TMAFP-m1 with no inner cysteine bonds are both approximately a factor of two smaller than the persistence lengths of the proteins with the bonds.

This also shows that the flexural persistence length in the direction studied of the SBAFP without inner cysteine bonds is even further from the value for the SBAFP-m1. This demonstrates that it must be due to other structural differences in the two proteins that contribute to the differences in their persistence lengths.

Another interesting result is that in general we see symmetry in the twisting directions for each protein, regardless of the natural twist direction of the protein. We also see the greatest asymmetry in the persistence lengths in the two bending directions for RiAFP, whose side-lengths also exhibit the most extreme asymmetry of all the proteins studied.

\subsection{Young's moduli}

\begin{figure}
	\center
	\includegraphics[width=0.5\textwidth]{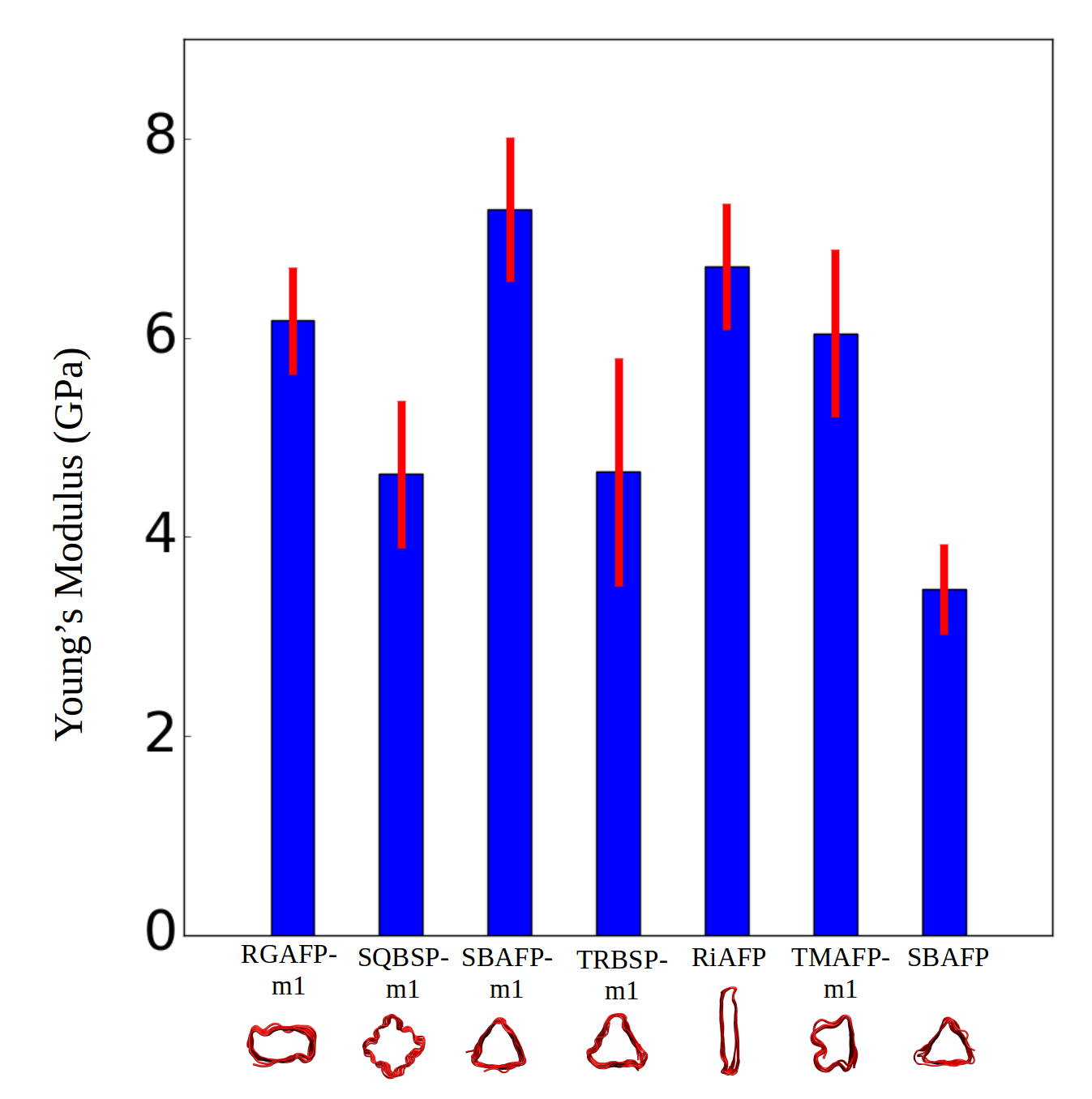}
	\caption{The average and standard deviation of the Young's modulus for each protein is shown.}
	\label{youngs_modulus}
\end{figure}

We studied the tensile strength of the proteins by measuring their Young's moduli. To do this for each protein, we pulled the center of mass of the last turn along its helical axis at a constant rate of 1 nm/ns until all the hydrogen bonds between the last and second-to-last turn were broken. We then analyzed the force-displacement data before breakage of the secondary structure. Umbrella sampling of the pulling trajectories was not done, due to the very small displacements achieved during pulling. This small displacement could not be properly sampled. Therefore the data shown in Fig. \ref{youngs_modulus} is from the non-equilibrium pulling simulations alone. 

The Young's modulus was calculated as $Y=kL_{0}/A$, where $L_{0}$ is the initial length of the protein, $A$ is its cross-sectional area, and $k$ is in this case the slope of the force-displacement curve derived from the pulling simulation. As mentioned by Solar and Buehler \cite{solar_2014}, it is important to note that small changes in the cross-sectional area used can result in significant changes in the Young's modulus. We included all residues in our area calculations, and so expect that our values for the Young's moduli are on the lower end of the potential range \cite{Knowles_2011}. (See Supplemental Information for more details on how we determined cross-sectional areas.)

Our results show that SBAFP-m1 and RiAFP both have relatively high Young's moduli, similar to each other within the error bars. SBAFP (the wild-type SBAFP) has the lowest, illustrating that its internal disulfide bonds do not provide tensile stiffening. TMAFP-m1, on the other hand, has a high Young's modulus (also comparable to SBAFP-m1 and RiAFP within the error bars) despite having just one beta sheet face per protein. In this case the internal disulfide bridges are oriented across the protein, back to front, and this may play a role in enhancing the tensile stability. 

\subsection{Packing fraction}

We analyzed the fraction of inward-facing residue volume to total residue volume to see whether it correlated with any strength measures (persistence length or Young's modulus). 

\begin{figure}
	\center
	\includegraphics[width=0.5\textwidth]{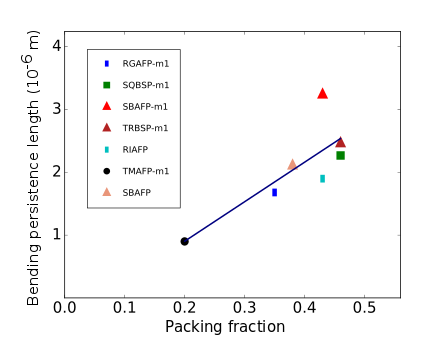}
	\caption{The relationship between bending persistence length and packing fraction for each protein is shown. The larger of the two average bending persistence lengths was chosen as the persistence length for each protein. The packing fraction is the ratio of inward-facing amino acid residues to total amino acid residues. Three-sided proteins are represented by triangles in shades of red, two-sided proteins are represented as rectangles in shades of blue, the four-sided protein is represented by a green square, and the ``one-sided'' protein is represented by an circle. The line of best fit is also shown, which has an $R^{2}$ value of 0.64.}
	\label{bending_PL_packing_fraction}
\end{figure}

\begin{figure}
	\center
	\includegraphics[width=0.5\textwidth]{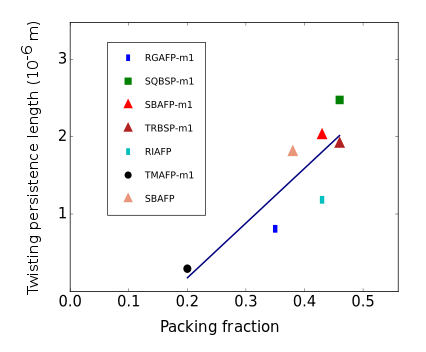}
	\caption{The relationship between twisting persistence length and packing fraction for each protein is shown. The larger of the two average twisting persistence lengths was chosen as the persistence length for each protein. The packing fraction is the ratio of inward-facing amino acid residues to total amino acid residues. Three-sided proteins are represented by triangles in shades of red, two-sided proteins are represented as rectangles in shades of blue, the four-sided protein is represented by a green square, and the ``one-sided'' protein is represented by a circle. The line of best fit is also shown, which has an $R^{2}$ value of 0.72.}
	\label{twisting_PL_packing_fraction}
\end{figure}

To do this, we compared the higher of the two flexural persistence lengths against the packing fraction for each protein, as well as the higher of the two torsional persistence lengths against the packing fraction. The results are plotted in Fig. \ref{bending_PL_packing_fraction} and Fig. \ref{twisting_PL_packing_fraction}. (For a table of packing fraction values and information on how they were calculated, see Supporting Information.)

We can see that a higher packing fraction does generally correlate to a higher persistence length, in both bending and twisting directions. The proteins with overall highest persistence lengths are SBAFP-m1, SQBSP-m1 and TRBSP-m1. These are also the proteins with the highest packing fractions. RiAFP and SBAFP are in the middle in both cases, and RGAFP-m1 and TMAFP-m1 have the lowest in both cases. In particular TMAFP-m1 clearly has the lowest packing fraction as well as lowest persistence lengths. 

To quantify the correlation, we computed linear least-squares regressions on the data plotted in Fig. \ref{bending_PL_packing_fraction} and Fig. \ref{twisting_PL_packing_fraction}. The fit line for flexural persistence length versus packing fraction is given by P$_{flex}(p) = 6.3p-0.4$, where $p$ is the packing fraction, with a coefficient of determination of $R^{2} = 0.64$. The line for the torsional persistence length vs packing fraction is given by P$_{twist}(p) = 7.1p-1.2$, with $R^{2} = 0.72$.

Comparisons of Fig. \ref{bending_PL_packing_fraction} and Fig. \ref{twisting_PL_packing_fraction} with Fig. \ref{youngs_modulus} show that the packing fraction does not correlate well with Young's modulus. This is most obvious for TMAFP-m1, which has a large Young's modulus, but small persistence lengths and packing fraction, demonstrating that while it is relatively weak under bending and twisting, its tensile stiffness is high.

\subsection{Hydrogen bonds}

We also examined whether the number of hydrogen bonds per turn correlated with persistence length or Young's modulus. For each protein we used VMD \cite{humphrey_96} to measure the number of hydrogen bonds in the protein over a 50 ps simulation. We took the average count over this period and divided it by the number of turns in the protein. (For details on distance and angle criteria for hydrogen bonds, and a table of values of hydrogen bonds per turn for each protein, see Supporting Information.)

\begin{figure}
	\center
	\includegraphics[width=0.5\textwidth]{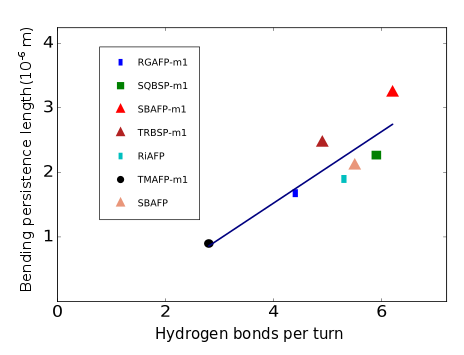}
	\caption{The relationship between bending persistence length and average number of hydrogen bonds per turn for each protein is shown. The larger of the two average bending persistence lengths was chosen as the persistence length for each protein. Three-sided proteins are represented by triangles in shades of red, two-sided proteins are represented as rectangles in shades of blue, the four-sided protein is represented by a green square, and the ``one-sided'' protein is represented by a circle. The line of best fit is also shown, which has an $R^{2}$ value of 0.77.}
	\label{bending_PL_hbonds}
\end{figure}

\begin{figure}
	\center
	\includegraphics[width=0.5\textwidth]{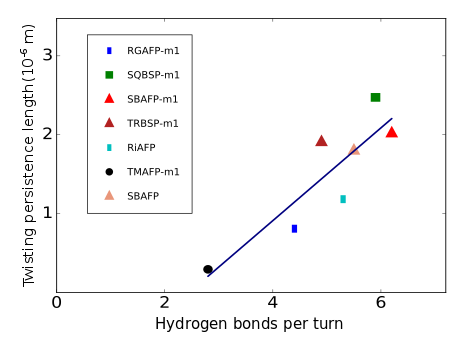}
	\caption{The relationship between twisting persistence length and average number of hydrogen bonds per turn for each protein is shown. The larger of the two average twisting persistence lengths was chosen as the persistence length for each protein. Three-sided proteins are represented by triangles in shades of red, two-sided proteins are represented as rectangles in shades of blue, the four-sided protein is represented by a green square, and the ``one-sided'' protein is represented by a circle. The line of best fit is also shown, which has an $R^{2}$ value of 0.77.}
	\label{twisting_PL_hbonds}
\end{figure}

The estimates for numbers of hydrogen bonds per turn are plotted against the largest bending and twisting persistence lengths for each protein and shown in Fig. \ref{bending_PL_hbonds} and Fig. \ref{twisting_PL_hbonds}, respectively.

From the plots, we can see that the average number of hydrogen bonds per turn does correlate well with both bending and twisting persistence length. RGAFP-m1 and TMAFP-m1 have the lowest numbers of hydrogen bonds per turn and have the lowest persistence lengths. SBAFP-m1 and SQBSP-m1 have the highest numbers of hydrogen bonds per turn and are among those with the highest persistence lengths. TRBSP-m1, on the other hand, is among the proteins with a lower numbers of hydrogen bonds per turn, and yet has generally high persistence lengths.

As we did in the analysis of the packing fraction, we calculated linear least-squares regressions for the data plotted in Fig. \ref{bending_PL_hbonds} and Fig. \ref{twisting_PL_hbonds}. For the flexural persistence length versus number of hydrogen bonds per turn, the fit line is given by P$_{flex}(h)$ = $0.6h-0.7$, where $h$ is the number of hydrogen bonds per turn, with a coefficient of determination of $R^{2} = 0.77$. For the torsional persistence length versus number of hydrogen bonds per turn, the fit line is given by P$_{twist}(h)$ = $0.6h-1.4$, with $R^{2} = 0.77$ as well.
 
We conclude that in addition to the packing fraction, the average number of hydrogen bonds per turn is a good predictor of persistence length magnitude.

One might also expect that the numbers of hydrogen bonds per turn to correlate well with quantities like Young's modulus, since hydrogen bonds are what maintain the rod-like structures of the proteins along their helical axes. This is clearly not the case, particularly for RGAFP-m1 and TMAFP-m1. We find that more hydrogen bonds does not equate to higher elastic stiffness.

\subsection{Pulling rates}

It should be noted that faster pulling rates have been shown to result in higher rupture forces \cite{gautieri_2008, ackbarow_2007}, and therefore larger Young's moduli, and that this may be due to the differences in the way hydrogen bonds are broken at different rates of pulling. Ackbarow \cite{ackbarow_2007} argues that at faster pulling rates, parallel hydrogen bonds between turns may break simultaneously in sets of three, necessitating larger applied forces, while at slower pulling rates, the bonds break sequentially, requiring less force to completely rupture the turn from its neighbor. 

We have seen this behavior as well, and since our pulling rate of 1 nm/ns is in the ``fast'' regime discussed in the literature (and is indeed quite fast compared to physical experiments), we propose that the breaking forces achieved may be higher than the forces necessary to break the secondary structure of these proteins. Regardless, the qualitative differences and trends in results between proteins and pulling directions, including strengths of proteins relative to each other, still remain true.

\section{Conclusions}

In conclusion we have performed a number of simulations of bending, twisting, and pulling on a variety of beta-solenoid proteins. For the bending and twist simulations, we have measured comparable values of the moduli from PMF and regression of force-displacement curves, but we can provide more easily quantifiable and tighter error bars from the latter. The resultant bend/twist persistence lengths are in the 1-3 micron scale with the three-sided SBAFP-m1 stiffest against bend and the 4-sided SQBSP-m1 stiffest against twist.  We find the strongest Young's modulus for the SBAFP-m1, followed closely by the two-sided RIAFP, RGAFP, and ``one-sided'' TMAFP-m1.  Although the internal intralayer disulfide bonds for the SBAFP wild-type did not confer strength compared to the SBAFP-m1 (where the cysteines are replaced by serines), the high Young's modulus of the TMAFP-m1 may reflect a reinforcement effect from internal disulfide bridges which in this case link the front and back side of the protein. The Young's moduli are all in the range of 3-7 GPa for these beta solenoid proteins, squarely in the range of such mechanically strong biomaterials as spider silk\cite{silk_youngs_modulus}. In addition, we found positive correlations between packing fraction and persistence length, as well as between the number of hydrogen bonds per turn and persistence length, for both bending and twisting directions.

\section{Acknowledgment}

The authors thank the UC Davis Office of Research RISE Program and the U.S. National Science Foundation for its support through the DMR-1207624 grant. In addition, we thank Michael Toney and his research group for useful conversations and insights.

\pagebreak

\bibliography{ref} 

\pagebreak

\section{Supporting Information}

\subsection{Validation of the beam model}

\begin{figure}[h]
	\center
	\includegraphics[width=0.6\textwidth]{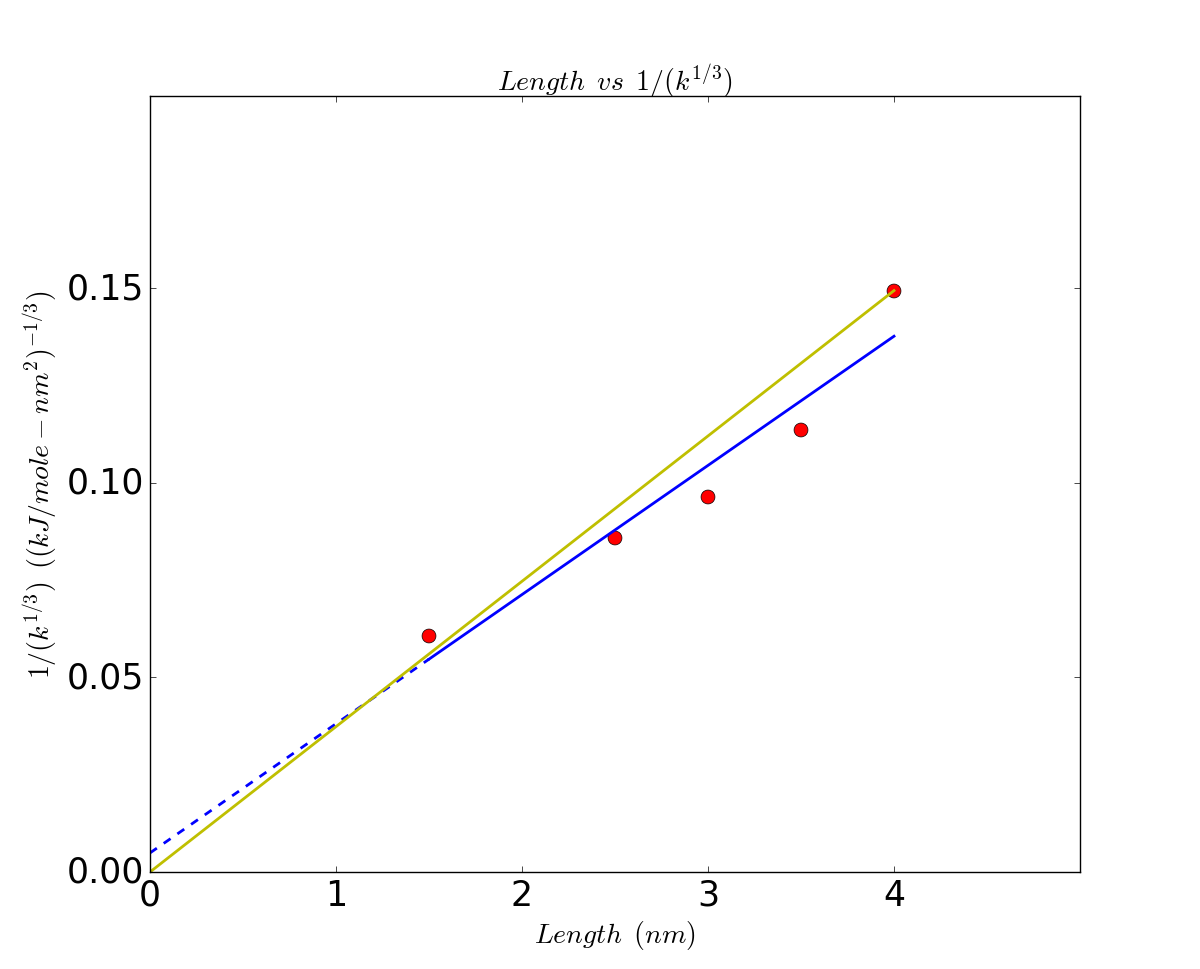}
	\caption{The relationship between one over the cube root of the spring constant and length for pulling of RiAFP in one of its bending directions is shown. The red dots are from simulation data. The yellow line is a fit between the original length of the protein and the origin. The blue line is a fit to the data points. The dashed blue line is the extrapolation of this fit to show the small, but non-zero, y intercept.}
	\label{riafp_var_length}
\end{figure}

It is not obvious that simple beam theory can accurately describe the behavior of our beta-solenoid proteins. As a check, we examined the relationship between spring constant and protein length. Based on the model, $L=Ck^{-1/3}$, where C is a constant ($C=(3(EI))^{1/3}$) and $L$ is the length of the protein from the fixed end to the end at which the force is being applied. We tested this relationship by varying the pulling location along the length of the protein and determining the spring constant to be the slope of the resulting force vs. displacement data. In effect, we roughly estimated the spring constants for different lengths of the protein. We then plotted length $L$ versus $k^{-1/3}$. An example is shown in Fig. \ref{riafp_var_length}. We saw generally linear behavior for all the proteins and directions tested, even at small lengths.

\subsection{Packing fraction calculation}

To calculate the packing fraction for each protein, we first determined whether each amino acid residue was inward-facing or outward-facing in its turn. We did this by first defining each turn as a path connecting the c-alpha atoms of the amino acid residues of that turn. For each residue in that turn, we calculated the average location of all its atoms. If that location lay inside the path defined by the backbone c-alphas, the residue was counted as being inward-facing. If instead it lay outside, it was counted as outward-facing. Once these were tabulated, the total volume of the inward-facing residues were calculated and divided by the total volume of all the residues in the protein. This ratio was taken as the packing fraction. The packing fraction value for each protein is listed in Table \ref{packing_fraction_table}.

\begin{table}
  \begin{center}
    \begin{tabular}{|c || c |} 
      \hline
      Protein & Packing fraction \\
      \hline\hline
      SQBSP-m1 & 0.46 \\ 
      \hline
      TRBSP-m1 & 0.46  \\
      \hline
      SBAFP-m1 & 0.43 \\
      \hline
      RiAFP & 0.43 \\
      \hline
      SBAFP & 0.38 \\  
      \hline
      RGAFP-m1 & 0.35 \\  
      \hline
      TMAFP-m1 & 0.20 \\  
      \hline
    \end{tabular}
      \caption{The packing fraction (fraction of inward-facing amino acid reside volume to total amino acid residue volume) for each protein is shown in descending order.}
      \label{packing_fraction_table}
  \end{center}
\end{table}

\subsection{Values of hydrogen bonds per turn}

In addition to the packing fraction, we investigated the relationship between numbers of hydrogen bonds per turn and persistence length. We first used the Hydrogen Bonds extension in VMD \cite{humphrey_96} to determine the total number of hydrogen bonds in the protein at each frame of a 50 ps simulation trajectory. The donor-acceptor distance used was 3.0 Angstroms and the angle cutoff was 20 degrees. Then we took the mean number of total hydrogen bonds over the trajectory and divided it by the number of turns in the protein.  

The values for average numbers of hydrogen bonds per turn for each protein are listed in Table \ref{hydrogen_bonds_table}.

\begin{table}
  \begin{center}
    \begin{tabular}{|c || c |} 
      \hline
      Protein & Hydrogen bonds per turn \\
      \hline\hline
      SBAFP-m1 & 6.2 \\
      \hline
      SQBSP-m1 & 5.9 \\ 
      \hline
      SBAFP & 5.5 \\  
      \hline
      RiAFP & 5.3 \\
      \hline
      TRBSP-m1 & 4.9  \\
      \hline
      RGAFP-m1 & 4.4 \\  
      \hline
      TMAFP-m1 & 2.8 \\  
      \hline
    \end{tabular}
      \caption{The average number of hydrogen bonds per turn in each protein are shown, in descending order. Each value was computed by taking the total hydrogen bonds in the protein (computed as the average over a 50 ps simulation) and dividing it by the number of total turns.}
      \label{hydrogen_bonds_table}
  \end{center}
\end{table}

\subsection{Beta content}

\begin{figure}
	\center
	\includegraphics[width=0.6\textwidth]{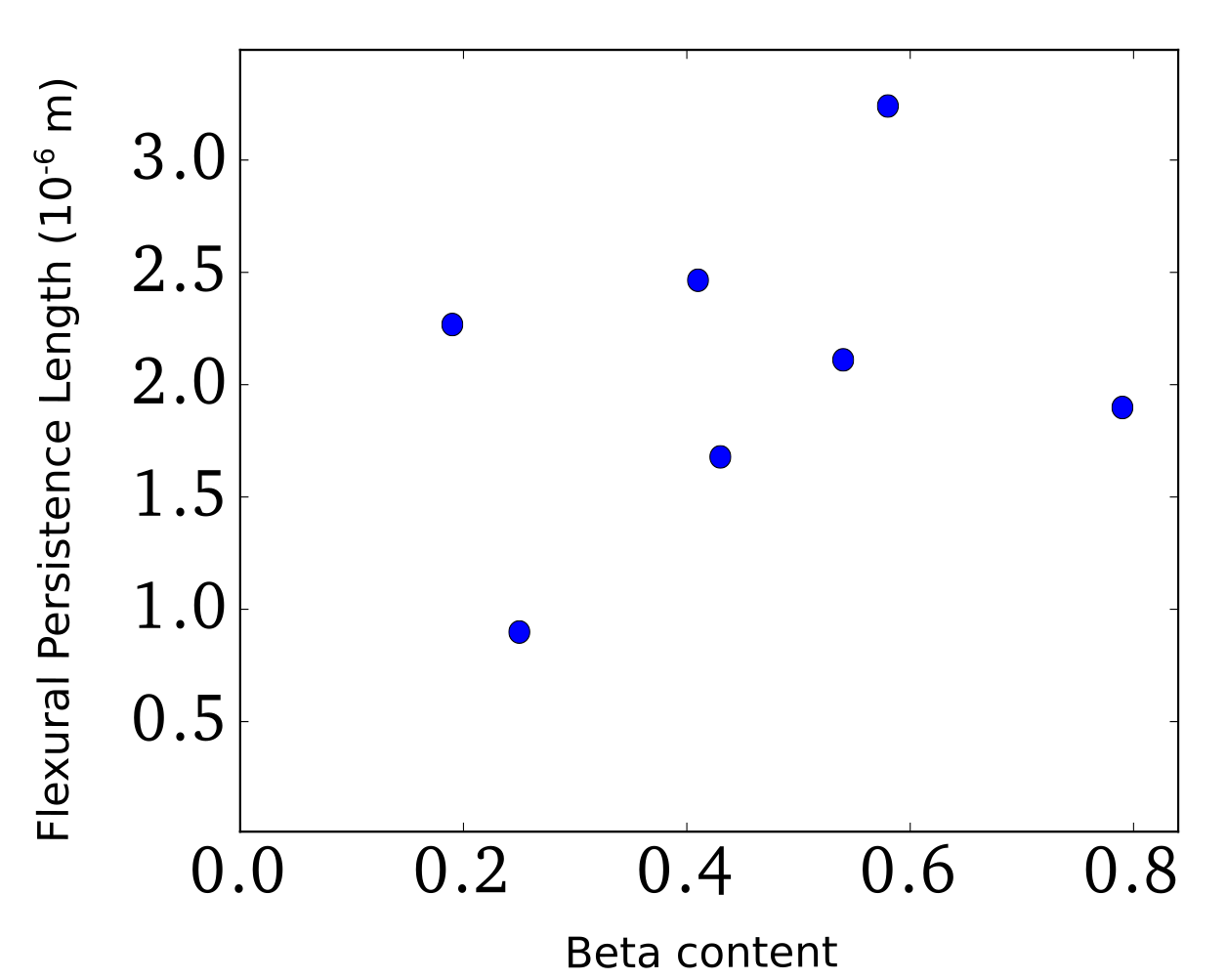}
	\caption{Flexural persistence length versus percent beta secondary structure.}
	\label{bending_PL_beta_content}
\end{figure}

\begin{figure}
	\center
	\includegraphics[width=0.6\textwidth]{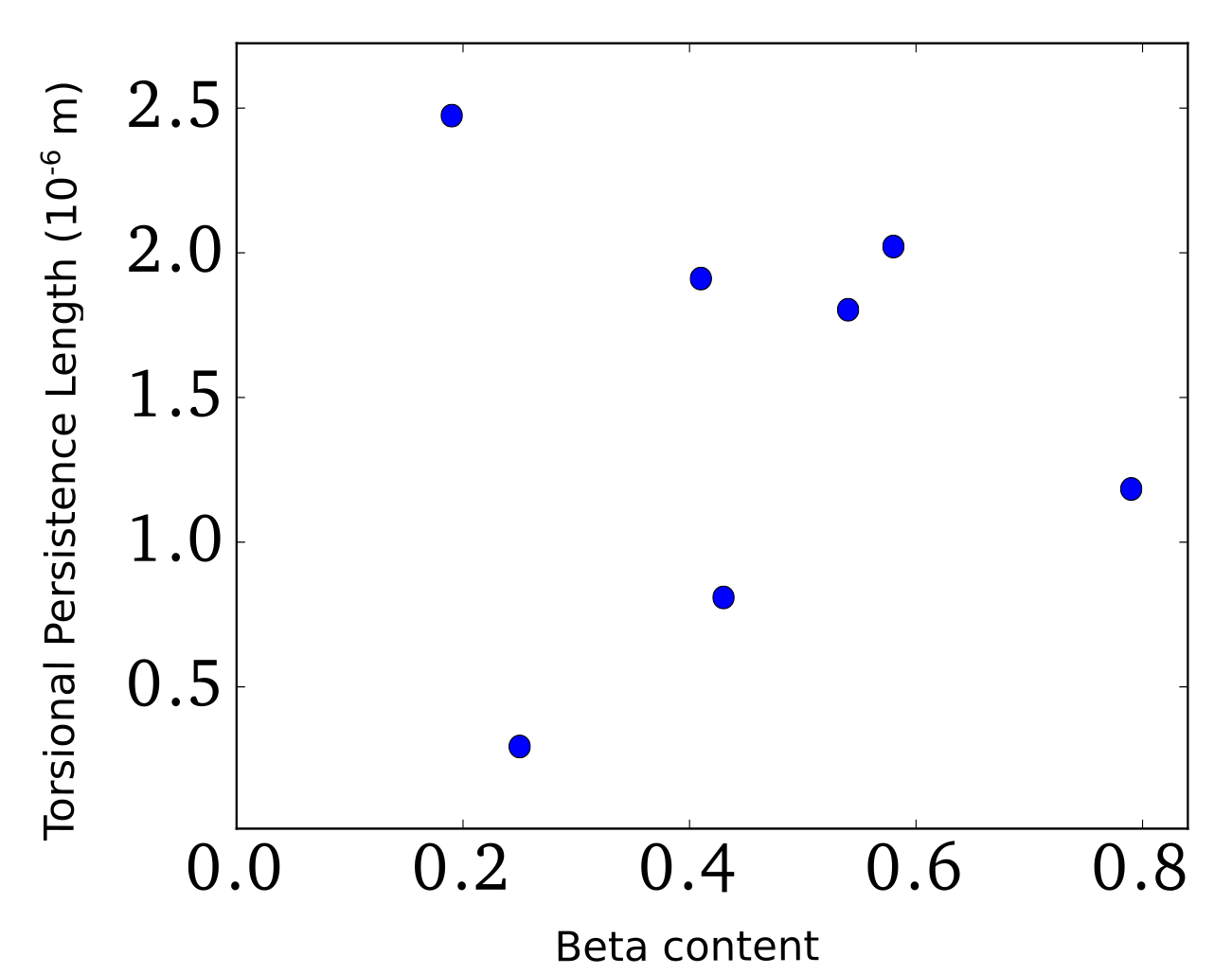}
	\caption{Torsional persistence length versus percent beta secondary structure.}
	\label{twisting_PL_beta_content}
\end{figure}

We also checked whether the beta content of each protein (beta sheets and beta bridges) correlated with persistence length. To do this, we used DSSP \cite{kabsch_1983} to calculate the amount of beta content in each protein. By this we mean the fraction of residues involved in beta strand or beta bridge secondary structures to the total number of residues. These values can be seen in Table \ref{beta_content_table} and the relationship between beta content and persistence length can be seen in Fig. \ref{bending_PL_beta_content} and Fig. \ref{twisting_PL_beta_content}.

\begin{table}
  \begin{center}
    \begin{tabular}{|c || c |} 
      \hline
      Protein & Beta content \\
      \hline\hline
      RiAFP & 0.79 \\
      \hline
      SBAFP-m1 & 0.58 \\
      \hline
      SBAFP & 0.54 \\  
      \hline
      RGAFP-m1 & 0.43 \\  
      \hline
       TRBSP-m1 & 0.41  \\
      \hline
      TMAFP-m1 & 0.25 \\  
      \hline
      SQBSP-m1 &  0.19 \\ 
      \hline
    \end{tabular}
      \caption{}
      \label{beta_content_table}
  \end{center}
\end{table}

From the plots, we can see that higher persistence length does not positively or negatively correlate with higher beta content.

\subsection{Cross-sectional area and density calculations}

The cross-sectional areas and densities of the proteins were calculated using a grid-sampling approach. Each protein was aligned so that its helical axis was along the x-axis. Each atom of each turn was represented as a sphere defined by its van der Waals radius. Each turn was then projected onto the y-z plane, so that it was now represented by a collection of overlapping circles. The turn was bounded by a box and this box was then divided into a grid with spacing of approximately 0.03 nm in each direction. This grid was sampled and points lying inside circles were counted as part of the turn's area. The total area of the turn of the protein was determined as the ratio of points lying inside circles to total grid points, multiplied by the total area of the box. To get a measure of the overall cross-sectional area of a protein, the areas of each turn of the protein were averaged.

\begin{figure}
	\center
	\includegraphics[width=0.6\textwidth]{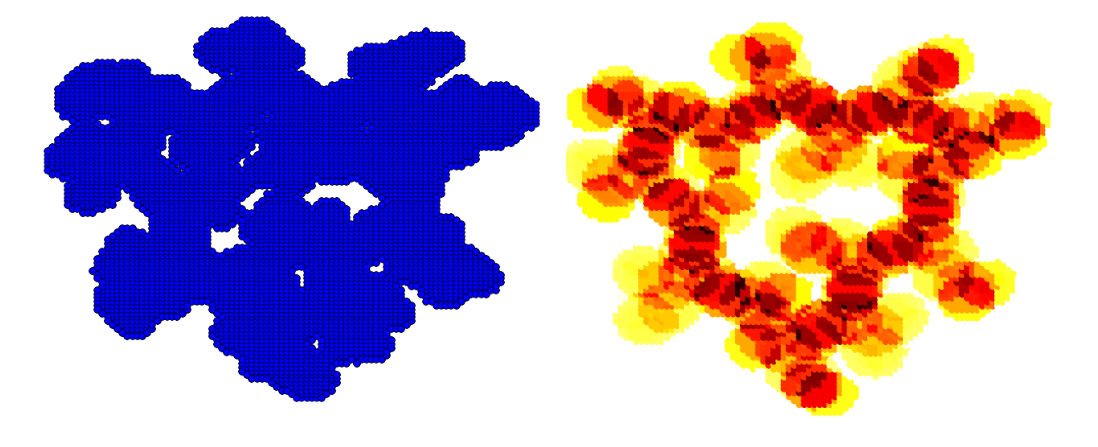}
	\caption{The total area (left) and density map (right) of the eighth turn of SBAFP-m1 is shown. Both were created using Python and Matplotlib by implementing a grid-sampling technique. In the case of the density map, the grid points were weighted by the area densities of the van der Waals circles representing the atoms in the turn.}
	\label{sbafpm1_area}
\end{figure}

The density was determined by giving a weight to each grid point lying inside a circle, based on the atom's weight and radius. This weight was essentially an area density, since it was calculated as the atom's mass divided by the area of the projected van der Waals sphere representing the atom. Points that lay inside multiple circles were weighted as the sum of the weights of each circle.

Examples of both the full cross-sectional area and density map are shown in Fig. \ref{sbafpm1_area} for one turn of SBAFP-m1.

\subsection{Protein Sequences}

\begin{table}
\begin{tabular}{c c c c c c c c c c }
ASP & LEU & SER & ILE & VAL & ASP & LEU & ARG & GLY & ALA \\ 
VAL & LEU & GLU & ASN & ILE & ASN & LEU & SER & GLY & ALA \\ 
ILE & LEU & HIS & GLY & ALA & MET & LEU & ASP & GLU & ALA \\ 
ASN & LEU & GLN & GLN & ALA & ASN & LEU & SER & ARG & ALA \\ 
ASP & LEU & SER & GLY & ALA & THR & LEU & ASN & GLY & ALA \\ 
ASP & LEU & ARG & GLY & ALA & ASN & LEU & SER & LYS & ALA \\ 
ASP & LEU & SER & ASP & ALA & ILE & LEU & ASP & ASN & ALA \\ 
ILE & LEU & GLU & GLY & ALA & ILE & LEU & ASP & GLU & ALA \\ 
VAL & LEU & ASN & GLN & ALA & ASN & LEU & LYS & ALA & ALA \\ 
ASN & LEU & GLU & GLN & ALA & ILE & LEU & SER & HIS & ALA \\ 
ASN & LEU & ARG & GLU & ALA & ASP & LEU & SER & GLU & ALA \\ 
ASN & LEU & GLU & ALA & ALA & ASP & LEU & SER & GLY & ALA \\ 
ASP & LEU & ALA & ILE & ALA & ASP & LEU & HIS & GLN & ALA \\ 
ASN & LEU & HIS & GLN & ALA & ALA & LEU & GLU & ARG & ALA \\ 
\end{tabular}
\caption{Sequence of SQBSP-m1.} 
\label{tab:seq_SQBSP-m1}
\end{table}

\begin{table}
\begin{tabular}{c c c c c c c c c c }
ALA & SER & ARG & ILE & THR & ASN & SER & GLN & ILE & VAL \\ 
LYS & SER & GLU & ALA & THR & ASN & SER & ASP & ILE & ASN \\ 
ASN & SER & GLN & LEU & VAL & ASP & SER & ILE & SER & THR \\ 
ARG & SER & GLN & TYR & SER & ASP & ALA & ASN & VAL & LYS \\ 
LYS & SER & VAL & THR & THR & ASP & SER & ASN & ILE & ASP \\ 
LYS & SER & GLN & VAL & TYR & LEU & THR & THR & SER & THR \\ 
GLY & SER & GLN & TYR & ASN & GLY & ILE & TYR & ILE & ARG \\ 
SER & SER & ASP & THR & THR & GLY & SER & GLU & ILE & SER \\ 
GLY & SER & SER & ILE & SER & THR & SER & ARG & ILE & THR \\ 
ASN & SER & ARG & ILE & THR & ASN & SER & GLN & ILE & VAL \\ 
LYS & SER & GLU & ALA & THR & ASN & SER & ASP & ILE & ASN \\ 
ASN & SER & GLN & LEU & VAL & ASP & SER & ILE & SER & THR \\ 
ARG & SER & GLN & TYR & SER & ASP & ALA & ASN & VAL & LYS \\ 
LYS & SER & VAL & THR & THR & ASP & SER & ASN & ILE & ASP \\ 
LYS & SER & GLN & VAL & TYR & LEU & THR & THR & SER & THR \\ 
GLY & SER & GLN & TYR & ASN & GLY & ILE & TYR & ILE & ARG \\ 
SER & SER & ASP & THR & THR & GLY & SER & GLU & ILE & SER \\ 
GLY & SER & SER & ILE & SER & THR & SER & ARG & ILE & THR \\ 
\end{tabular}
\caption{Sequence of TRBSP-m1.} 
\label{tab:seq_TRBSP-m1}
\end{table}

\begin{table}
\begin{tabular}{c c c c c c c c c c }
ASN & ASP & ILE & ASP & GLY & THR & ASN & ASN & GLU & VAL \\ 
ASP & GLY & SER & GLU & ASN & VAL & LEU & ALA & GLY & ASN \\ 
ASP & ASN & THR & VAL & SER & GLY & ASP & ASN & ASN & SER \\ 
VAL & SER & GLY & SER & ASN & ASN & THR & VAL & SER & GLY \\ 
ASN & ASP & ASN & THR & VAL & THR & GLY & SER & ASN & HIS \\ 
VAL & VAL & SER & GLY & THR & ASN & HIS & ILE & VAL & THR \\ 
ASP & ASN & ASN & ASN & ASN & VAL & SER & GLY & ASN & ASP \\ 
ASN & ASN & VAL & SER & GLY & SER & PHE & HIS & THR & VAL \\ 
SER & GLY & GLY & HIS & ASN & THR & VAL & SER & GLY & SER \\ 
ASN & ASN & THR & VAL & SER & GLY & LYS & ARG & HIS & ARG \\ 
VAL & GLN & GLY & THR & ASN & ASN & ARG & VAL & THR & ASP \\ 
\end{tabular}
\caption{Sequence of RGAFP-m1.} 
\label{tab:seq_RGAFP-m1}
\end{table}

\begin{table}
\begin{tabular}{c c c c c c c c c c }
GLY & VAL & GLU & ILE & GLY & GLU & GLY & THR & VAL & LEU \\ 
LYS & SER & GLY & VAL & VAL & VAL & ASN & GLY & GLY & THR \\ 
LYS & ILE & GLY & ARG & ASP & ASN & GLU & ILE & TYR & GLN \\ 
GLY & ALA & SER & ILE & GLY & GLY & GLY & VAL & GLU & ILE \\ 
GLY & ASP & ARG & ASN & ARG & ILE & ARG & GLU & SER & VAL \\ 
THR & ILE & GLY & GLY & GLY & GLY & VAL & VAL & GLY & SER \\ 
ASP & ASN & LEU & LEU & MET & ILE & ASN & ALA & GLY & ILE \\ 
ALA & GLY & ASP & CYS & THR & VAL & GLY & ASN & ARG & CYS \\ 
ILE & LEU & ALA & ASN & ASN & ALA & THR & LEU & ALA & GLY \\ 
GLY & VAL & GLU & ILE & GLY & GLU & GLY & THR & VAL & LEU \\ 
LYS & SER & GLY & VAL & VAL & VAL & ASN & GLY & GLY & THR \\ 
LYS & ILE & GLY & ARG & ASP & ASN & GLU & ILE & TYR & GLN \\ 
GLY & ALA & SER & ILE & GLY & GLY & GLY & VAL & GLU & ILE \\ 
GLY & ASP & ARG & ASN & ARG & ILE & ARG & GLU & SER & VAL \\ 
THR & ILE & GLY & GLY & GLY & GLY & VAL & VAL & GLY & SER \\ 
ASP & ASN & LEU & LEU & MET & ILE & ASN & ALA & GLY & ILE \\ 
ALA & GLY & ASP & CYS & THR & VAL & GLY & ASN & ARG & CYS \\ 
ILE & LEU & ALA & ASN & ASN & ALA & THR & LEU & ALA & GLY \\ 
\end{tabular}
\caption{Sequence of SBAFP-m1.} 
\label{tab:seq_SBAFP-m1}
\end{table}

\begin{table}
\begin{tabular}{c c c c c c c c c c}
ALA & SER & ARG & ALA & GLU & ALA & ARG & GLY & GLU & ALA \\ 
MET & ALA & GLU & GLY & HIS & SER & ARG & GLY & CYS & ALA \\
THR & SER & HIS & ALA & ASN & ALA & THR & GLY & HIS & ALA \\ 
ASP & ALA & ARG & SER & MET & SER & GLU & GLY & ASN & ALA \\
GLU & ALA & TYR & THR & GLU & ALA & LYS & GLY & THR & ALA \\ 
MET & ALA & THR & SER & GLU & ALA & SER & GLY & GLU & ALA \\
ARG & ALA & GLN & THR & ASN & ALA & ASP & GLY & ARG & ALA \\ 
HIS & SER & SER & SER & ARG & THR & HIS & GLY & ARG & ALA \\
ASP & SER & THR & ALA & SER & ALA & LYS & GLY & GLU & ALA \\ 
MET & ALA & GLU & GLY & THR & SER & ASP & GLY & ASP & ALA \\
LYS & SER & TYR & ALA & SER & ALA & ASP & GLY & ASN & ALA \\ 
CYS & ALA & LYS & SER & MET & SER & THR & GLY & HIS & ALA \\
ASP & ALA & THR & THR & ASN & ALA & HIS & GLY & THR & ALA \\ 
MET & ALA & ASP & SER & ASN & ALA & ILE & GLY & GLU & ALA \\
ARG & ALA & GLU & THR & ARG & ALA & GLU & GLY & ARG & ALA \\ 
GLU & SER & SER & SER & ASP & THR & ASP & GLY & CYS \\
\end{tabular}
\caption{Sequence of RiAFP.} 
\label{tab:seq_RiAFP}
\end{table}

\begin{table}
\begin{tabular}{c c c c c c c c c c c c}
GLY & ALA & CYS & THR & GLY & CYS & GLY & ASN & CYS & PRO & ASN & ALA \\
VAL & THR & CYS & THR & ASN & SER & GLN & HIS & CYS & VAL & LYS & ALA \\
ASN & THR & CYS & THR & GLY & SER & THR & ASP & CYS & ASN & THR & ALA \\
GLN & THR & CYS & THR & ASN & SER & LYS & ASP & CYS & PHE & GLU & ALA \\
ASN & THR & CYS & THR & ASP & SER & THR & ASN & CYS & TYR & LYS & ALA \\
THR & ALA & CYS & THR & ASN & SER & SER & GLY & CYS & PRO \\
\end{tabular}
\caption{Sequence of TMAFP.} 
\label{tab:seq_TMAFP}
\end{table}

\begin{table}
\begin{tabular}{c c c c c c c c c c c c}
GLY & THR & CYS & VAL & ASN & THR & ASN & SER & GLN & ILE & THR & ALA \\ 
ASN & SER & GLN & CYS & VAL & LYS & SER & THR & ALA & THR & ASN & CYS \\ 
TYR & ILE & ASP & ASN & SER & GLN & LEU & VAL & ASP & THR & SER & ILE \\ 
CYS & THR & ARG & SER & GLN & TYR & SER & ASP & ALA & ASN & VAL & LYS \\ 
LYS & SER & VAL & THR & THR & ASP & CYS & ASN & ILE & ASP & LYS & SER \\ 
GLN & VAL & TYR & LEU & THR & THR & CYS & THR & GLY & SER & GLN & TYR \\ 
ASN & GLY & ILE & TYR & ILE & ARG & SER & SER & THR & THR & THR & GLY \\ 
THR & SER & ILE & SER & GLY & PRO & GLY & CYS & SER & ILE & SER & THR \\ 
CYS & THR & ILE & THR \\
\end{tabular}
\caption{Sequence of SBAFP.} 
\label{tab:seq_SBAFP}
\end{table}

\end{document}